\documentclass[a4paper,twocolumn,prl]{revtex4-2}%

\usepackage[latin9]{inputenc}
\usepackage{amsmath}
\usepackage{amssymb}
\usepackage{graphicx}
\usepackage{comment}
\usepackage{mathrsfs}
\usepackage{amsfonts}
\usepackage{float}%
\usepackage{multirow}
\usepackage{color}
\providecommand{\U}[1]{\protect\rule{.1in}{.1in}}
\makeatletter

\DeclareFontEncoding{LGR}{}{}
\DeclareTextSymbol{\~}{LGR}{126}
\@ifundefined{textcolor}{}
{
\definecolor{BLACK}{gray}{0}
\definecolor{WHITE}{gray}{1}
\definecolor{RED}{rgb}{1,0,0}
\definecolor{GREEN}{rgb}{0,1,0}
\definecolor{BLUE}{rgb}{0,0,1}
\definecolor{CYAN}{cmyk}{1,0,0,0}
\definecolor{MAGENTA}{cmyk}{0,1,0,0}
\definecolor{YELLOW}{cmyk}{0,0,1,0}
}

\makeatother

\begin{document}
\title{   Giant \textcolor{blue}{Apparent} Flexoelectricity in Semiconductors Driven by Insulator-to-metal Transition }
\author{Y.-X. Wang$^{1,2}$}
\thanks{YXW and JGL contributed equally to this work.}
\author{J.-G. Li$^{1,2}$}
\thanks{YXW and JGL contributed equally to this work.}
\author{G. Seifert$^{3}$}
\author{K. Chang$^{4,5,6}$}
\thanks{Corresponding author}
\email[]{kchang@semi.ac.cn}
\author{D.-B. Zhang$^{2}$}
\thanks{Corresponding author}
\email[]{dbzhang@bnu.edu.cn}
\affiliation{$^{1}$College of Nuclear Science and Technology, Beijing Normal University, Beijing 100875, P.R. China}
\affiliation{$^{2}$Department of Physics, Beijing Normal University, Beijing 100875, P.R. China}
\affiliation{$^{3}$Theoretische Chemie, Technische Universitat Dresden, D-01062 Dresden, Germany}
\affiliation{$^{4}$SKLSM, Institute of Semiconductors, Chinese Academy of Sciences, P.O. Box 912, Beijing 100083, China}
\affiliation{$^{5}$CAS Center for Excellence in Topological Quantum Computation, University of Chinese Academy of Sciences, Beijing 100190, China}
\affiliation{$^{6}$Beijing Academy of Quantum Information Sciences, Beijing 100193, China}
\begin{abstract}
We elucidate the flexoelectricity of materials in the high strain gradient regime, of which the underlying mechanism is less understood. By using the generalized Bloch theorem, we uncover a strong \textcolor{blue}{flexoelectric-like} effect in bent thinfilms of Si and Ge due to a high strain gradient-induced insulator-to-metal transition. We show that an unusual type-II band alignment is formed between the compressed and elongated sides of the bent film, resulting in a spatial separation of electron and hole. Therefore, upon the insulator-to-metal transition, electrons transfer from the compressed side to the elongated side to reach the thermodynamic equilibrium, leading to pronounced polarization along the film thickness dimension. The obtained transverse flexoelectric coefficients are unexpectedly high, with a quadratic dependence on the film thickness. This new mechanism is extendable to other semiconductor materials with moderate energy gaps. Our findings have important implications for the future \textcolor{blue}{applications} of flexoelectricity in semiconductor materials.
\end{abstract}

\maketitle

\textcolor{blue}{The   flexoelectricity describes the electric polarization of materials in response to  mechanical strain gradients~\cite{review1,review2,review3,review4,addstengel1,addstengel2}. It is believed to exist not only in dielectric materials, but also in metals~\cite{revise1,revise2,revise3}}.  Recent experimental advances have revealed unexpectedly strong  \textcolor{blue}{flexoelectric-like effects} in semiconductors with high strain gradients~\cite{wangzl,pv,mos2,high9,high11}. This renews the interest in functionalizing semiconductors toward excellent electromechanical applications. Mechanically, the presence of a high strain gradient indicates a complex deformation of the structure, where inhomogeneous strain patterns are involved~\cite{high2,high1,high3,high7,high8}. Due to the severe structural deformations, the structure-electronic properties relation may assume a new physical mechanism, which is unfortunately less investigated. Theoretically, atomistic approaches that can address the complex strain patterns in a facile way are the proper methods   to address the problem.

\textcolor{blue}{A high strain gradient can be realized in nano-sized objects or by introducing local deformations to bulk materials~\cite{high9,revise4}. Under the high strain gradient condition, the resulting flexoelectric response may adopt a nonlinear behavior due to the significant contribution of the higher order terms in the flexoelectric effect~\cite{revise5}. Although several analyses based on the ionic chain model of polar systems suggest that such nonlinearity may cause a reduction in the polarization strength~\cite{high6,high4}, many experimental investigations  have revealed strong electric polarizations upon the presence of high strain gradients not only in systems with high flexoelectric coefficients~\cite{high5,high7,high8} but also in those with low flexoelectric coefficients~\cite{pv,wangzl}. For example, for bulk Si, the first-principles calculation  obtains  flexoelectric coefficient value of $\sim1$~nCm$^{-1}$~\cite{hong3}. However, recent experimentation in nanoindentation has reported the flexoelectric coefficient to be as high as $\sim78$~nCm$^{-1}$~\cite{wangzl}, where the strain gradient approaches $10^6$m$^{-1}$. This suggests that the formation of electric polarization and the associated flexoelectric-like effect may have  different origins. }

In this work, we reveal that strong \textcolor{blue}{flexoelectric-like} effects can be realized through an insulator-to-metal transition in bent semiconductor thinfilms. The high strain gradient is induced by a simple bending deformation. We uncover a novel type-II band alignment between the compressed and elongated sides of the bent film, realizing a spatial separation of electron and hole. As such, upon  band gap closure, electrons immigrate from the compressed side to the elongated side to reach the thermodynamic equilibrium. Such a charge transfer leads to strong polarization in the film thickness dimension.

We demonstrate this idea by performing quantum mechanical simulations of bent thinfilms of Si and Ge. These simulations are enabled  by employing the generalized Bloch theorem~\cite{gbt1,gbt2} in which the bent motif is described with  rotational symmetry~\cite{gbt5,rotate,rotate2,rotate3,rotate4}.   Our simulations successfully illustrate the formation of the type-II band alignment and the insulator-to-metal phase transition at a critical strain gradient. With this, the charge polarization and the related flexoelectric properties are next estimated. The obtained \textcolor{blue}{effective} transverse flexoelectric coefficient displays a quadratic dependence on the film thickness \textcolor{blue}{and becomes comparable to the reported experimental value  at a relatively small thickness less than $100$ nm}.

\begin{figure}[tb]
\includegraphics[width=0.8\columnwidth]{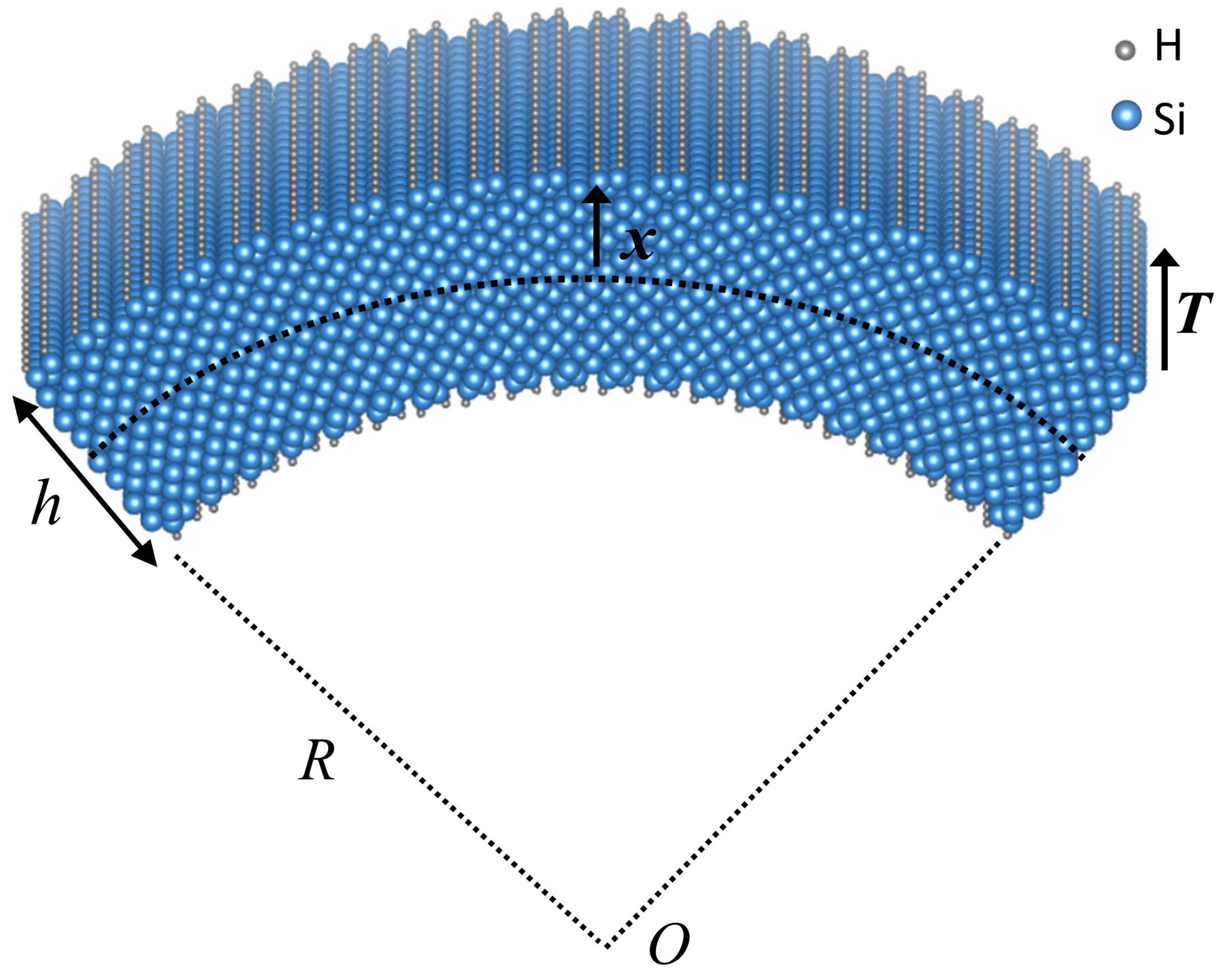}
\caption{ Bent motif of the (100) film  with a principle curvature of $1/R$. ${\bf T}$ is the translation vector that is invariant against bending. $h$ denotes the film thickness.  $-h/2<x<h/2$ along the thickness dimension measures the distance from the neutral surface (dashed curve) of the bent film. }%
\label{film}%
\end{figure}

The thinfilms with  thickness $h$ in our simulations are (100)-oriented. Using the generalized Bloch theorem, pure bending with a principal curvature of $1/R$ can be applied to the film, Fig.~\ref{film}, where compressional and tensile strains are distributed antisymmetrically on the opposite sides of the middle neutral surface. As such, no lateral strain is invoked and the resulting strain gradient is simply $g=1/R$, being uniform throughout the film thickness dimension. \textcolor{blue}{This pure bending condition allows us to precisely control the transverse strain gradient~\cite{Stengel1,rotate}.}  We  investigate the electronic properties of bent Si (100) thinfilms with thicknesses $h$ ranging from \textcolor{blue}{30} to 200~nm and a bent Ge (100) thinfilm with a 10 nm thickness using the generalized Bloch theorem coupled with  density-functional tight-binding~\cite{dftb1,dftb2,dftb3}. \textcolor{blue}{The atomic structures of films are fully relaxed.} For these films, the  dangling bonds at the surface regions are passivated with H atoms to diminish the surplus surface states. The atomic populations are evaluated using the Mulliken charge analysis. \textcolor{blue}{More details of computational settings are provided in Section I of the Supplemental material}.

Fig.~\ref{gap} displays the band structure of (a) the stress-free (100) Si film  with wavenumber $k$ in the [001] direction and accordingly, the band structures  for the bent film at strain gradients,\textcolor{blue}{ (b) $g=4.8~\mu$m$^{-1}$  and (c) $g=8.0~\mu$m$^{-1}$}, \textcolor{blue}{see Fig. S4 and Fig. S5 of the Supplemental material for more information of the band structures}. For the stress-free film, the band structure features a direct band gap at $k=0$, Fig.~\ref{gap}(a). Bending has a strong influence on the electronic spectrum of Si film, which is manifested mostly in the significant reduction of  the fundamental band gap, Fig.~\ref{gap}(b). Interestingly, at greater strain gradients, the band gap can be closed, Fig.~\ref{gap}(c). This indicates an insulator-to-metal phase transition due to bending. We note that the onset of the band gap closure is at a strain gradient \textcolor{blue}{ of $g=7.8~\mu$m$^{-1}$ for this thickness}.  A similar result of band gap closure is also obtained in the bent Ge film, see \textcolor{blue}{Fig. S8 of} the Supplemental material.

\begin{figure}[tb]
\includegraphics[width=1\columnwidth]{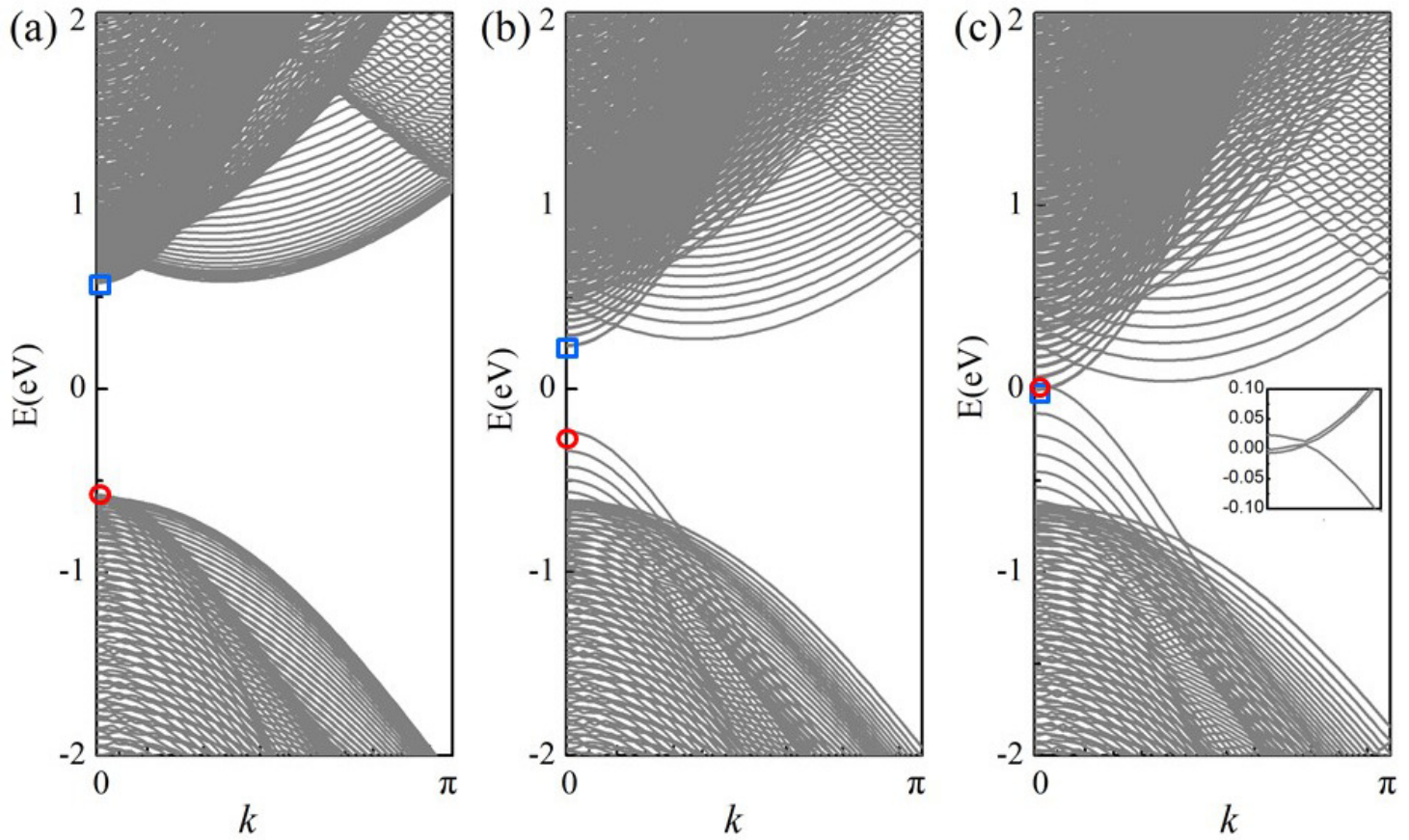}
\caption{Comparison of electronic band structures of (a) the stress-free ($g=0$) and bent (100) Si films with (b) \textcolor{blue}{ $g=4.8~\mu$m$^{-1}$ and (c) $g=8.0~\mu$m$^{-1}$. The film is 30 nm thick.} The Fermi energy is set at zero. Opened squares and circles locate the CBM state and VBM state at $k=0$, respectively. In (c), the inset magnifies the band structure around the Fermi energy.}%
\label{gap}%
\end{figure}

\begin{figure}[tb]
\includegraphics[width=1\columnwidth]{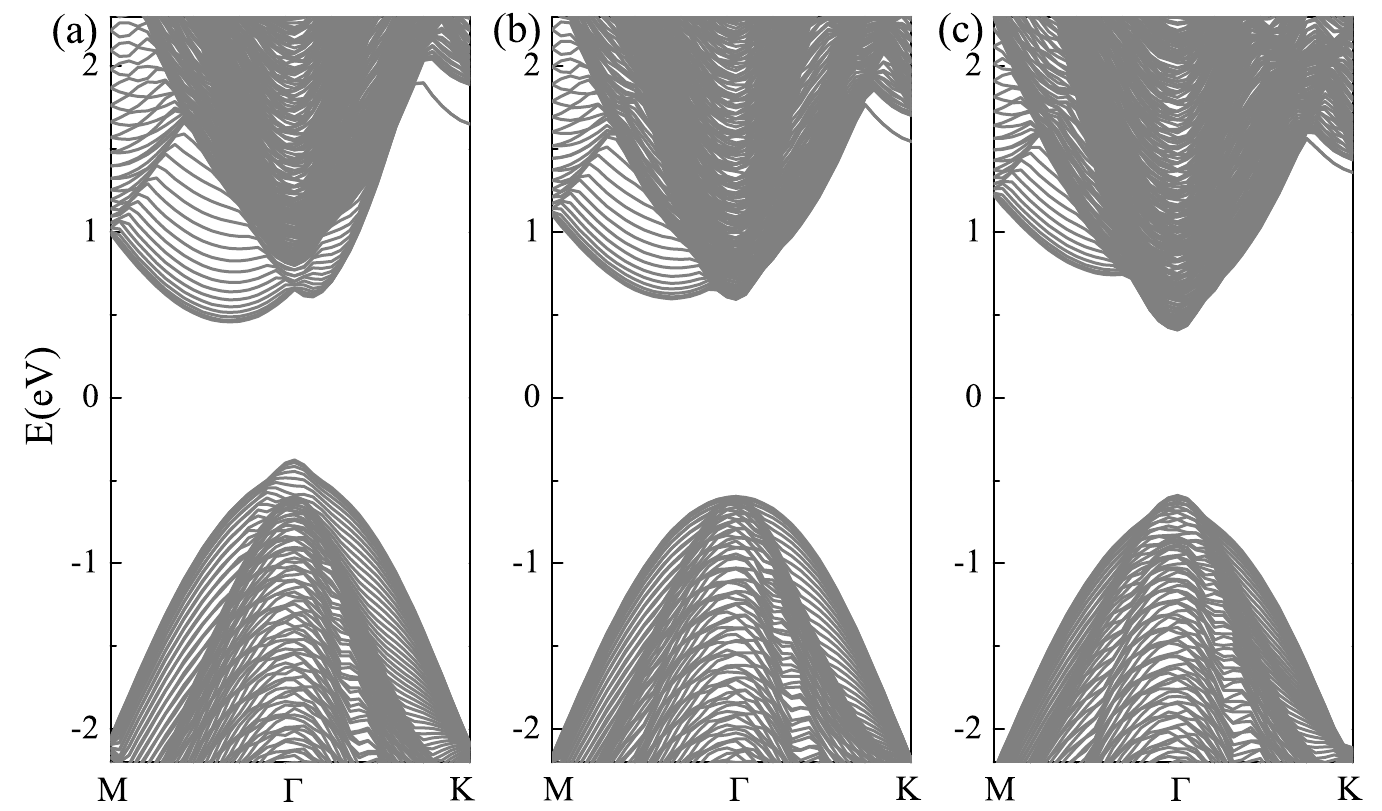}
\caption{ \textcolor{blue}{Electronic band structures of a 30 nm thick Si film under different uniaxial strains along the [100] crystallographic direction: (a) compressional strain  $\epsilon=-3\%$, (b) stress-free $\epsilon=0$, and (c) tensile strain $\epsilon=3\%$.}  }%
\label{strain}%
\end{figure}

\begin{figure}[tb]
\includegraphics[width=1\columnwidth]{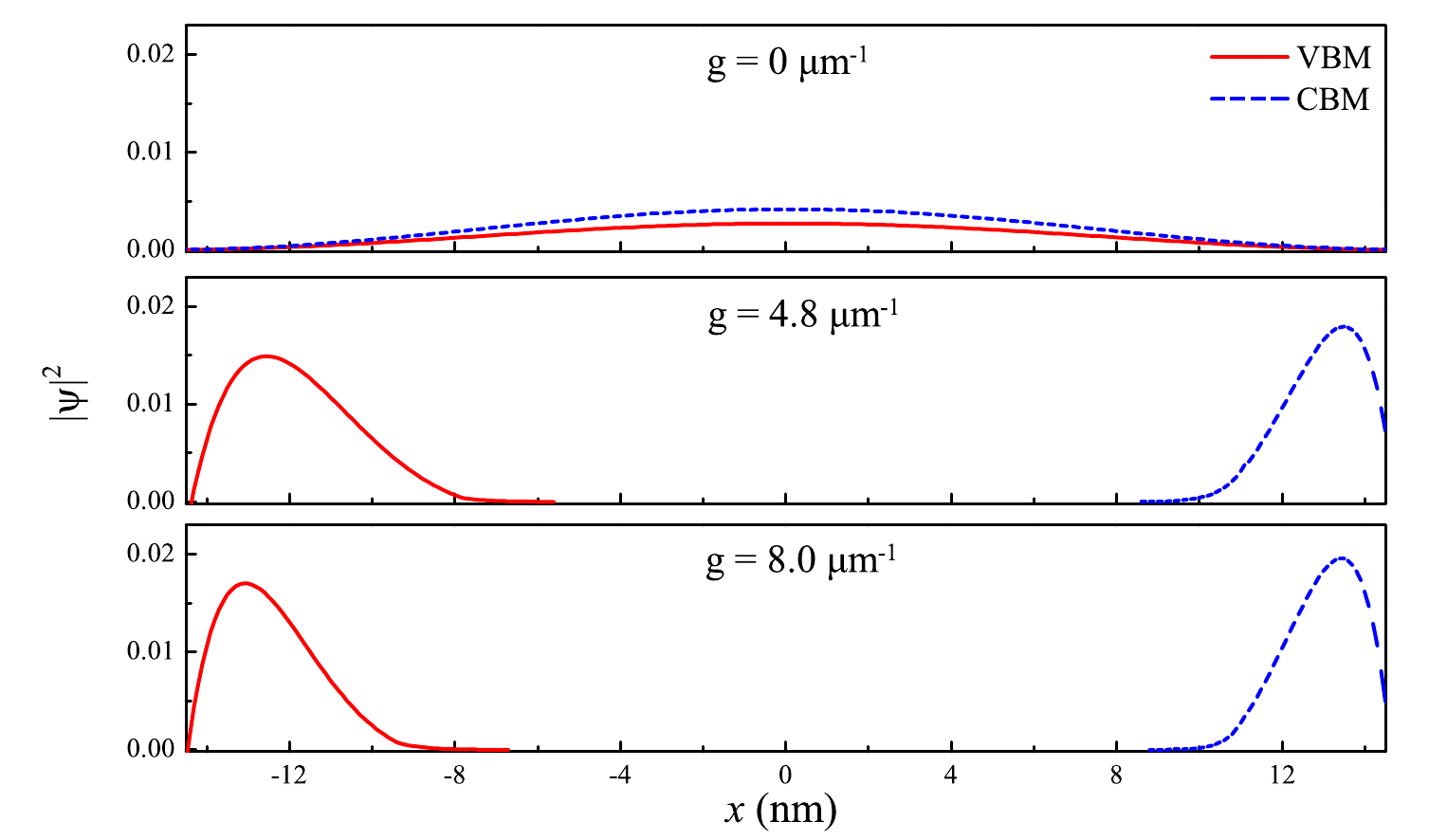}
\caption{ \textcolor{blue}{Cross section of the wavefunction ($|\Psi|^2$)} in the film thickness dimension, $x$, of  the electronic states whose locations in the energy spectrum are indicated by   opened squares and circles  in Fig.~\ref{gap}, for the stress-free film ($g=0$) and bent films at two strain gradients \textcolor{blue}{($g=4.8~\mu$m$^{-1}$  and $g=8.0~\mu$m$^{-1}$)}.  }%
\label{wave}%
\end{figure}

We indicate that the closure of the band gap originates from the formation of an effective type-II band alignment in the bent film.  To illustrate the mechanism, we study the variation in the band structure of strained Si films.  Uniaxial strain is applied to films along the [100] crystallographic direction. \textcolor{blue}{For a physically intuitive illustration of the relative band alignment between films with different strains, we determine all band energies with respect to the vacuum energy level that is common for different films, see Section X of the Supplemental material for an interpretation to this special consideration}. Fig.~\ref{strain} shows the band structures of the film with different uniaxial strains. \textcolor{blue}{We note that the band shifts with strains are not rigid and} focus on the frontier conduction bands (CBs) and valence bands (VBs) \textcolor{blue}{ around the Fermi level}. Compared to the stress-free film, the \textcolor{blue}{frontier} CB states adopt downshifts for both the compressed and elongated films, but the shift of the elongated film is much greater than that of the compressed film. Consequently, the \textcolor{blue}{frontier} CB states of the elongated film are lower in energy than those of the compressed film. On the other hand, the variation in frontier VB states assumes a reverse trend. Compared to the stress-free film, the \textcolor{blue}{frontier} VB states exhibit upshifts for  both the compressed and elongated films but the shift of the compressed film is much greater than that of the elongated film. Thus, the \textcolor{blue}{frontier} VB states of the compressed film are higher in energy than those of the elongated film. We have also carried out first-principles calculations~\cite{vasp1,vasp2,vasp3} using VASP~\cite{vasp} to confirm these trends; see the report in the Supplemental Material.  From this analysis, we conclude that the frontier VB and CB states of the bent film  are localized, residing around the compressed side and the elongated side, respectively.  As a demonstration, the wavefunctions of the conduction band minimum (CBM) state and valence band maximum (VBM) state at $k=0$ are visualized. These states are highlighted by scatters in Fig.~\ref{gap}. Fig.~\ref{wave} displays the spatial distributions of the wavefunctions of the CBM and VBM states along the film thickness dimension of the 30 nm thick Si film. For the stress-free film,  both CBM and VBM states are extensive over the whole film thickness. Under bending, both states become localized with the CBM (VBM) state residing on the elongated (compressed) side of the bent film, which persists even after the insulator-metal phase transition.

\begin{figure}[t]
\includegraphics[width=1\columnwidth]{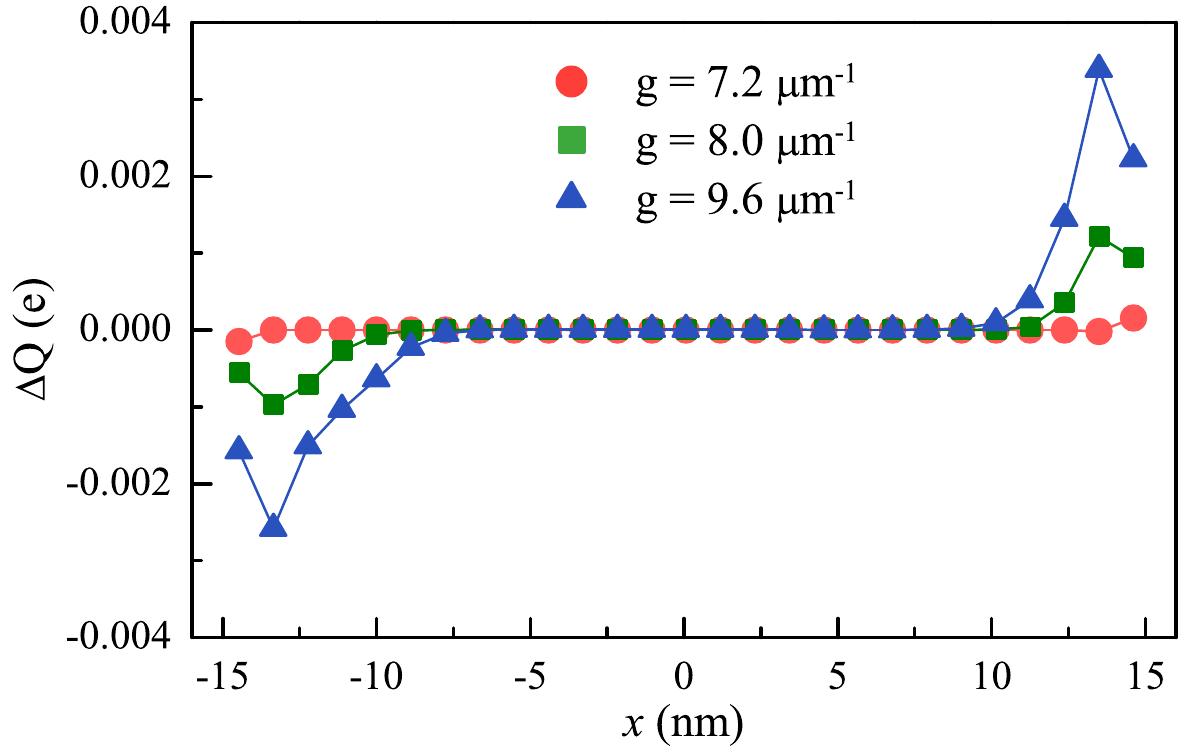}
\caption{ Distribution of net charges per atom along  the film thickness dimension, $x$, of the \textcolor{blue}{30~nm} thick bent films at different strain gradients, $g$.  }%
\label{charge}%
\end{figure}

The spatial separation of the frontier VB and CB states (or equivalently, the separation of electron and hole) in the bent film mimics the special electronic configuration of the type-II band alignment that is usually formed in heterostructures~\cite{typeii1,typeii2,typeii3}.  Due to such spatial separation of electron and hole, the immigration of electronic charges from the compressed side to the elongated side of the bent film is triggered when the band gap closes. Indeed,  upon  gap closure, the energy of CB states becomes lower than that of VB states, such as the VBM and CBM states around  $k=0$ shown in Fig.~\ref{gap}(c). Consequently, the CB states are filled with electrons, while the VB states are empty. This hints a charge transfer from the compressed side to the elongated side along the film thickness dimension since the film is charge neutral as a whole.

Based on this analysis, we next evaluate the variation in atomic charges due to bending. Notice that the atomic population $Q$ consists  of  the contributions from all the occupied states. Using atomic orbitals as basis, we calculate $Q$ via  Mulliken charge analysis~\cite{gbt5}. With this, the variation in the atomic charge due to bending is obtained as,
\begin{equation}
\label{netq}
\triangle Q=Q_{\text{bent}}(x)-Q_{\text{stress-free}}(x),
\end{equation}
where $Q_{\text{bent}}$  refers to the population of the atom located at position $x$  of the bent film and $Q_{\text{stress-free}}$ is the same but for the stress-free film. Fig.~\ref{charge} displays the obtained $\triangle Q$ of the \textcolor{blue}{30} nm thick Si films  with different  strain gradients.  Before the insulator-to-metal transition, at a strain gradient \textcolor{blue}{ $g=7.2~\mu$m$^{-1}$,}  $\triangle Q$ is relatively large only for those atoms around the film surfaces where $\triangle Q<0$ for atoms on the compressive side and $\triangle Q>0$ for the atoms on the elongated side. For other atoms inside the film, $\triangle Q$ is vanishingly small. These results indicate that the charge transfer along the film thickness is not important. Such a distribution of $\triangle Q$ is due to the redistribution of the bound charges in an insulator~\cite{Tagantsev,Resta1,hong2}.  Physically, this can be attributed to an emerged polarity of the Si-Si covalent bonds in the bent film~\cite{gbt5}. Upon the onset of the insulator-to-metal transition, $\triangle Q$ is pronouncedly large not only for the surface atoms but also for those atoms inside the film, indicating that the charge transfer along the film thickness is significant now. Accordingly, the resulting polarization is expected to be considerably strong. The insulator-to-metal transition-induced significant polarization for Ge is shown in Section VIII of the Supplemental Material. \textcolor{blue}{ Note that the contribution of surface flexoelectricity~\cite{Stengel2} to this polarization is not significant~\cite{surfacejap} and the impact of surface configurations on the polarization is also little as shown in Section IX of the Supplemental Material.}

\begin{figure}[t]
\includegraphics[width=1\columnwidth]{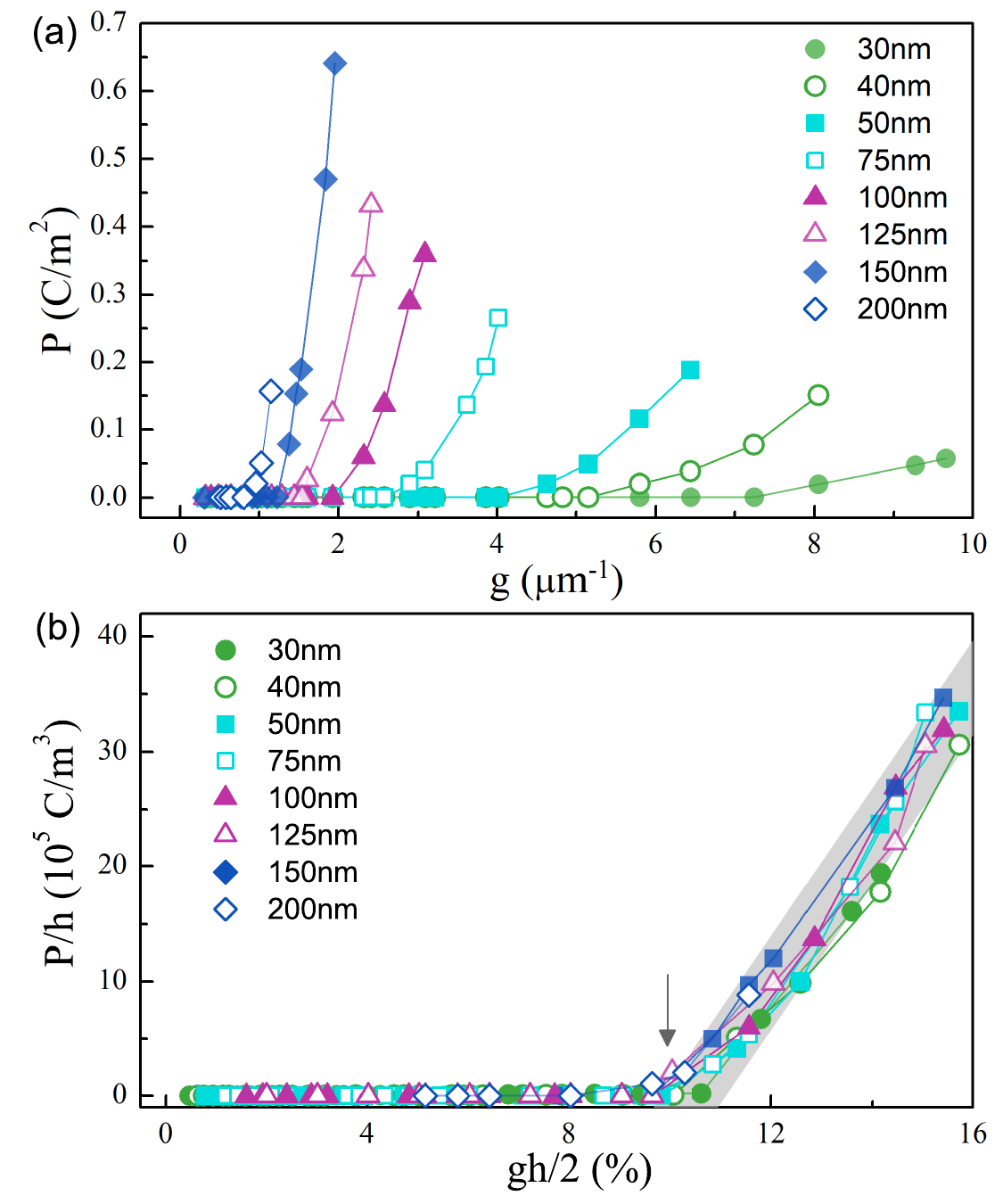}
\caption{ Electric polarization of the bent Si films with different thicknesses: (a) ${\bf P}$ versus \textcolor{blue}{$g$} and (b) ${\bf P}/h$ versus $gh/2$.  }%
\label{data}%
\end{figure}

With the atomic charge, the electric polarization is evaluated simply as,
\begin{equation} \label{polar}
{\bf P}=\frac{1}{\Omega}\int_{-h/2}^{h/2}\triangle Q(x)xdx,
\end{equation}
where, $\Omega=T_0^{2}h$ denotes the volume of the simulation cell used in the generalized Bloch theorem. The coordinate origin is at the position of the neutral surface, $x=0$. By symmetry, it is obvious that ${\bf P}\simeq0$ for the stress-free film. As such, ${\bf P}$ denotes the net polarization along the film thickness dimension induced by bending. The results of Si  films with different thicknesses are summarized in \textcolor{blue}{Fig.~\ref{data}(a). Important aspects can be revealed by studying the behavior of ${\bf P}/h$ versus $gh/2$, Fig.~\ref{data}(b). Note} that $gh/2$ equals the strain at the film surface on the elongated side in the bent film (The strain at the opposite surface is $-gh/2$) according to the pure bending condition as shown in Fig.~\ref{film}.  First,  the rapid increases in ${\bf P}$ due to the insulator-to-metal transition for all the considered thicknesses  start at a value of \textcolor{blue}{ $gh/2\simeq9.8\%$}, as guided by the downward arrow, Fig.~\ref{data}(a). The coincidence of this critical $gh/2$ value  for different Si films can be explained by the fact that the insulator-to-metal transition in a bent Si film is initiated by the band gap closure between the VB states around the surface on the compressed side and the CB states around the surface on the elongated side of the bent film, Fig.~\ref{gap}(c). Notice that the shifts of the VB and CB states are essentially driven by the strains on the opposite surfaces.  Because the size dependence of the electronic states on the film thickness is negligible, the strains needed for the insulator-to-metal transition are thus similar for different films. \textcolor{blue}{Note that this value of strains ($\simeq9.8\%$) is still within the elastic regime of Si~\cite{strains}}. Second, a collapse of the polarization curves is found when ${\bf P}$ is scaled by $h$, Fig.~\ref{data}(b). Such a scaling collapse provides insights into the thickness dependence property of the transverse flexoelectric coefficient and thus makes it convenient to derive  the  coefficient from the polarization data,  $\mu={\partial}{\bf P}/{\partial}g~{\simeq}~Dh^2$, where the constant \textcolor{blue}{$D=3.2{\times}10^{7}$ Cm$^{-3}$} is identified from the slope of the collapsed curves as guided by the gray bar. \textcolor{blue}{With $\mu$, it is also possible to derive the converse flexoelectric effect in the form of a force per unit area along the film thickness dimension,  \begin{equation}
f=(\varepsilon E'+\mu/R)Ek,
\end{equation}
where, the external electric field $E'=E(1-xk)$ is along the film thickness dimension. $k$ is a constant. $E$ represents the strength of $E'$, see the Section IV of the Supplemental material for more details.} The quadratic thickness dependence of the obtained transverse flexoelectric coefficient reflects the unique behavior of the electric polarization induced by the insulator-to-metal transition.

In summary, using the generalized Bloch theorem and taking Si and Ge as examples, we have shown that strong flexoelectric effects can be induced in bent semiconductor thinfilms due to an insulator-to-metal phase transition.  This result cannot be explained by the simple model of redistribution of bound charges, rather it is driven by electron transfer from the compressed side to the elongated side of the bent structure enabled by the insulator-to-metal transition. Attributed to this new mechanism, the obtained transverse flexoelectric coefficient exhibits a strong dependence on the film thickness.  Our results provide an interpretation to the strong flexoelectric effects found in the high strain gradient regime. Indeed, according to our calculation, the transverse flexoelectric coefficient $\mu=100$ nCm$^{-1}$ at a thickness of \textcolor{blue}{$h=56$ nm}, which is comparable to the experimental value~\cite{wangzl}. More importantly, according to this new mechanism, strong flexoelectric effects may be realized in many other semiconductors with moderate energy gaps,  such as those IV-VI components~\cite{pbte,pbx,viiv} and III-V~\cite{singh,harrison} components. Because the band gaps are small,  to induce the insulator-to-metal transition in these semiconductors, the needed structural deformation can be still within the mechanically elastic regime. These findings pave a new route to search for and explore strong flexoelectricity and associated novel applications in semiconductor materials.

DBZ thanks Su-Huai Wei, and Jiawang Hong for insightful suggestions. This work was supported by the MOST of China under Grant No. 2017Y
FA0303400, and by the NSFC under Grants Nos. 11874088, 12274035, and 11674022. D.-B.Z. was supported by the Fundamental Research Funds for the Central Universities. Computations were performed at the Beijing Computational Science Research Center and Beijing Normal University.

\bibliography{references}

\begin{thebibliography}{54}%
\makeatletter
\providecommand \@ifxundefined [1]{%
 \@ifx{#1\undefined}
}%
\providecommand \@ifnum [1]{%
 \ifnum #1\expandafter \@firstoftwo
 \else \expandafter \@secondoftwo
 \fi
}%
\providecommand \@ifx [1]{%
 \ifx #1\expandafter \@firstoftwo
 \else \expandafter \@secondoftwo
 \fi
}%
\providecommand \natexlab [1]{#1}%
\providecommand \enquote  [1]{``#1''}%
\providecommand \bibnamefont  [1]{#1}%
\providecommand \bibfnamefont [1]{#1}%
\providecommand \citenamefont [1]{#1}%
\providecommand \href@noop [0]{\@secondoftwo}%
\providecommand \href [0]{\begingroup \@sanitize@url \@href}%
\providecommand \@href[1]{\@@startlink{#1}\@@href}%
\providecommand \@@href[1]{\endgroup#1\@@endlink}%
\providecommand \@sanitize@url [0]{\catcode `\\12\catcode `\$12\catcode
  `\&12\catcode `\#12\catcode `\^12\catcode `\_12\catcode `\%12\relax}%
\providecommand \@@startlink[1]{}%
\providecommand \@@endlink[0]{}%
\providecommand \url  [0]{\begingroup\@sanitize@url \@url }%
\providecommand \@url [1]{\endgroup\@href {#1}{\urlprefix }}%
\providecommand \urlprefix  [0]{URL }%
\providecommand \Eprint [0]{\href }%
\providecommand \doibase [0]{https://doi.org/}%
\providecommand \selectlanguage [0]{\@gobble}%
\providecommand \bibinfo  [0]{\@secondoftwo}%
\providecommand \bibfield  [0]{\@secondoftwo}%
\providecommand \translation [1]{[#1]}%
\providecommand \BibitemOpen [0]{}%
\providecommand \bibitemStop [0]{}%
\providecommand \bibitemNoStop [0]{.\EOS\space}%
\providecommand \EOS [0]{\spacefactor3000\relax}%
\providecommand \BibitemShut  [1]{\csname bibitem#1\endcsname}%
\let\auto@bib@innerbib\@empty
\bibitem [{\citenamefont {Yudin}\ and\ \citenamefont
  {Tagantsev}(2013)}]{review1}%
  \BibitemOpen
  \bibfield  {author} {\bibinfo {author} {\bibfnamefont {P.~V.}\ \bibnamefont
  {Yudin}}\ and\ \bibinfo {author} {\bibfnamefont {A.~K.}\ \bibnamefont
  {Tagantsev}},\ }\bibfield  {title} {\bibinfo {title} {Fundamentals of
  flexoelectricity in solids},\ }\href@noop {} {\bibfield  {journal} {\bibinfo
  {journal} {Nanotechnology}\ }\textbf {\bibinfo {volume} {24}},\ \bibinfo
  {pages} {432001} (\bibinfo {year} {2013})}\BibitemShut {NoStop}%
\bibitem [{\citenamefont {Zubko}\ \emph {et~al.}(2013)\citenamefont {Zubko},
  \citenamefont {Catalan},\ and\ \citenamefont {Tagantsev}}]{review2}%
  \BibitemOpen
  \bibfield  {author} {\bibinfo {author} {\bibfnamefont {P.}~\bibnamefont
  {Zubko}}, \bibinfo {author} {\bibfnamefont {G.}~\bibnamefont {Catalan}},\
  and\ \bibinfo {author} {\bibfnamefont {A.~K.}\ \bibnamefont {Tagantsev}},\
  }\bibfield  {title} {\bibinfo {title} {Flexoelectric effect in solids},\
  }\href@noop {} {\bibfield  {journal} {\bibinfo  {journal} {Annu. Rev. Mater.
  Res.}\ }\textbf {\bibinfo {volume} {43}},\ \bibinfo {pages} {387} (\bibinfo
  {year} {2013})}\BibitemShut {NoStop}%
\bibitem [{\citenamefont {Wang}\ \emph
  {et~al.}(2019{\natexlab{a}})\citenamefont {Wang}, \citenamefont {Gu},
  \citenamefont {Zhang},\ and\ \citenamefont {Chen}}]{review3}%
  \BibitemOpen
  \bibfield  {author} {\bibinfo {author} {\bibfnamefont {B.}~\bibnamefont
  {Wang}}, \bibinfo {author} {\bibfnamefont {Y.}~\bibnamefont {Gu}}, \bibinfo
  {author} {\bibfnamefont {S.}~\bibnamefont {Zhang}},\ and\ \bibinfo {author}
  {\bibfnamefont {L.-Q.}\ \bibnamefont {Chen}},\ }\bibfield  {title} {\bibinfo
  {title} {Flexoelectricity in solids: Progress, challenges, and
  perspectives},\ }\href@noop {} {\bibfield  {journal} {\bibinfo  {journal}
  {Prog. Mater. Sci.}\ }\textbf {\bibinfo {volume} {106}},\ \bibinfo {pages}
  {100570} (\bibinfo {year} {2019}{\natexlab{a}})}\BibitemShut {NoStop}%
\bibitem [{\citenamefont {Deng}\ \emph {et~al.}(2020)\citenamefont {Deng},
  \citenamefont {Lv}, \citenamefont {Li}, \citenamefont {Tan}, \citenamefont
  {Liang},\ and\ \citenamefont {Shen}}]{review4}%
  \BibitemOpen
  \bibfield  {author} {\bibinfo {author} {\bibfnamefont {Q.}~\bibnamefont
  {Deng}}, \bibinfo {author} {\bibfnamefont {S.}~\bibnamefont {Lv}}, \bibinfo
  {author} {\bibfnamefont {Z.}~\bibnamefont {Li}}, \bibinfo {author}
  {\bibfnamefont {K.}~\bibnamefont {Tan}}, \bibinfo {author} {\bibfnamefont
  {X.}~\bibnamefont {Liang}},\ and\ \bibinfo {author} {\bibfnamefont
  {S.}~\bibnamefont {Shen}},\ }\bibfield  {title} {\bibinfo {title} {The impact
  of flexoelectricity on materials, devices, and physics},\ }\href@noop {}
  {\bibfield  {journal} {\bibinfo  {journal} {J. Appl. Phys.}\ }\textbf
  {\bibinfo {volume} {128}},\ \bibinfo {pages} {080902} (\bibinfo {year}
  {2020})}\BibitemShut {NoStop}%
\bibitem [{\citenamefont {Stengel}(2013{\natexlab{a}})}]{addstengel1}%
  \BibitemOpen
  \bibfield  {author} {\bibinfo {author} {\bibfnamefont {M.}~\bibnamefont
  {Stengel}},\ }\bibfield  {title} {\bibinfo {title} {Flexoelectricity from
  density-functional perturbation theory},\ }\href@noop {} {\bibfield
  {journal} {\bibinfo  {journal} {Phys. Rev. B}\ }\textbf {\bibinfo {volume}
  {88}},\ \bibinfo {pages} {174106} (\bibinfo {year}
  {2013}{\natexlab{a}})}\BibitemShut {NoStop}%
\bibitem [{\citenamefont {Dreyer}\ \emph {et~al.}(2018)\citenamefont {Dreyer},
  \citenamefont {Stengel},\ and\ \citenamefont {Vanderbilt}}]{addstengel2}%
  \BibitemOpen
  \bibfield  {author} {\bibinfo {author} {\bibfnamefont {C.~E.}\ \bibnamefont
  {Dreyer}}, \bibinfo {author} {\bibfnamefont {M.}~\bibnamefont {Stengel}},\
  and\ \bibinfo {author} {\bibfnamefont {D.}~\bibnamefont {Vanderbilt}},\
  }\bibfield  {title} {\bibinfo {title} {Current-density implementation for
  calculating flexoelectric coefficients},\ }\href@noop {} {\bibfield
  {journal} {\bibinfo  {journal} {Phys. Rev. B}\ }\textbf {\bibinfo {volume}
  {98}},\ \bibinfo {pages} {075153} (\bibinfo {year} {2018})}\BibitemShut
  {NoStop}%
\bibitem [{\citenamefont {Zabalo}\ and\ \citenamefont
  {Stengel}(2021)}]{revise1}%
  \BibitemOpen
  \bibfield  {author} {\bibinfo {author} {\bibfnamefont {A.}~\bibnamefont
  {Zabalo}}\ and\ \bibinfo {author} {\bibfnamefont {M.}~\bibnamefont
  {Stengel}},\ }\bibfield  {title} {\bibinfo {title} {Switching a polar metal
  via strain gradients},\ }\href@noop {} {\bibfield  {journal} {\bibinfo
  {journal} {Phys. Rev. Lett.}\ }\textbf {\bibinfo {volume} {126}},\ \bibinfo
  {pages} {127601} (\bibinfo {year} {2021})}\BibitemShut {NoStop}%
\bibitem [{\citenamefont {Yurkov}\ and\ \citenamefont {Yudin}(2021)}]{revise2}%
  \BibitemOpen
  \bibfield  {author} {\bibinfo {author} {\bibfnamefont {A.~S.}\ \bibnamefont
  {Yurkov}}\ and\ \bibinfo {author} {\bibfnamefont {P.~V.}\ \bibnamefont
  {Yudin}},\ }\bibfield  {title} {\bibinfo {title} {Flexoelectricity in
  metals},\ }\href@noop {} {\bibfield  {journal} {\bibinfo  {journal} {J. Appl.
  Phys.}\ }\textbf {\bibinfo {volume} {129}},\ \bibinfo {pages} {195108}
  (\bibinfo {year} {2021})}\BibitemShut {NoStop}%
\bibitem [{\citenamefont {Morozovska}\ \emph {et~al.}(2011)\citenamefont
  {Morozovska}, \citenamefont {Eliseev}, \citenamefont {Tagantsev},
  \citenamefont {Bravina}, \citenamefont {Chen},\ and\ \citenamefont
  {Kalinin}}]{revise3}%
  \BibitemOpen
  \bibfield  {author} {\bibinfo {author} {\bibfnamefont {A.~N.}\ \bibnamefont
  {Morozovska}}, \bibinfo {author} {\bibfnamefont {E.~A.}\ \bibnamefont
  {Eliseev}}, \bibinfo {author} {\bibfnamefont {A.~K.}\ \bibnamefont
  {Tagantsev}}, \bibinfo {author} {\bibfnamefont {S.~L.}\ \bibnamefont
  {Bravina}}, \bibinfo {author} {\bibfnamefont {L.-Q.}\ \bibnamefont {Chen}},\
  and\ \bibinfo {author} {\bibfnamefont {S.~V.}\ \bibnamefont {Kalinin}},\
  }\bibfield  {title} {\bibinfo {title} {Thermodynamics of electromechanically
  coupled mixed ionic-electronic conductors: Deformation potential, vegard
  strains, and flexoelectric effect},\ }\href@noop {} {\bibfield  {journal}
  {\bibinfo  {journal} {Phys. Rev. B}\ }\textbf {\bibinfo {volume} {83}},\
  \bibinfo {pages} {195313} (\bibinfo {year} {2011})}\BibitemShut {NoStop}%
\bibitem [{\citenamefont {Wang}\ \emph
  {et~al.}(2020{\natexlab{a}})\citenamefont {Wang}, \citenamefont {Liu},
  \citenamefont {Feng}, \citenamefont {Zhang}, \citenamefont {Zhu},
  \citenamefont {Zhai}, \citenamefont {Qin},\ and\ \citenamefont
  {Wang}}]{wangzl}%
  \BibitemOpen
  \bibfield  {author} {\bibinfo {author} {\bibfnamefont {L.}~\bibnamefont
  {Wang}}, \bibinfo {author} {\bibfnamefont {S.}~\bibnamefont {Liu}}, \bibinfo
  {author} {\bibfnamefont {X.}~\bibnamefont {Feng}}, \bibinfo {author}
  {\bibfnamefont {C.}~\bibnamefont {Zhang}}, \bibinfo {author} {\bibfnamefont
  {L.}~\bibnamefont {Zhu}}, \bibinfo {author} {\bibfnamefont {J.}~\bibnamefont
  {Zhai}}, \bibinfo {author} {\bibfnamefont {Y.}~\bibnamefont {Qin}},\ and\
  \bibinfo {author} {\bibfnamefont {Z.~L.}\ \bibnamefont {Wang}},\ }\bibfield
  {title} {\bibinfo {title} {Flexoelectronics of centrosymmetric
  semiconductors},\ }\href@noop {} {\bibfield  {journal} {\bibinfo  {journal}
  {Nature Nanotechnol.}\ }\textbf {\bibinfo {volume} {15}},\ \bibinfo {pages}
  {661} (\bibinfo {year} {2020}{\natexlab{a}})}\BibitemShut {NoStop}%
\bibitem [{\citenamefont {Yang}\ \emph {et~al.}(2018)\citenamefont {Yang},
  \citenamefont {Kim},\ and\ \citenamefont {Alexe}}]{pv}%
  \BibitemOpen
  \bibfield  {author} {\bibinfo {author} {\bibfnamefont {M.-M.}\ \bibnamefont
  {Yang}}, \bibinfo {author} {\bibfnamefont {D.~J.}\ \bibnamefont {Kim}},\ and\
  \bibinfo {author} {\bibfnamefont {M.}~\bibnamefont {Alexe}},\ }\bibfield
  {title} {\bibinfo {title} {Flexo-photovoltaic effect},\ }\href@noop {}
  {\bibfield  {journal} {\bibinfo  {journal} {Science}\ }\textbf {\bibinfo
  {volume} {360}},\ \bibinfo {pages} {904} (\bibinfo {year}
  {2018})}\BibitemShut {NoStop}%
\bibitem [{\citenamefont {Jiang}\ \emph {et~al.}(2021)\citenamefont {Jiang},
  \citenamefont {Chen}, \citenamefont {Hu}, \citenamefont {Xiang},
  \citenamefont {Zhang}, \citenamefont {Wang}, \citenamefont {Wang},\ and\
  \citenamefont {Shi}}]{mos2}%
  \BibitemOpen
  \bibfield  {author} {\bibinfo {author} {\bibfnamefont {J.}~\bibnamefont
  {Jiang}}, \bibinfo {author} {\bibfnamefont {Z.}~\bibnamefont {Chen}},
  \bibinfo {author} {\bibfnamefont {Y.}~\bibnamefont {Hu}}, \bibinfo {author}
  {\bibfnamefont {Y.}~\bibnamefont {Xiang}}, \bibinfo {author} {\bibfnamefont
  {L.}~\bibnamefont {Zhang}}, \bibinfo {author} {\bibfnamefont
  {Y.}~\bibnamefont {Wang}}, \bibinfo {author} {\bibfnamefont {G.-C.}\
  \bibnamefont {Wang}},\ and\ \bibinfo {author} {\bibfnamefont
  {J.}~\bibnamefont {Shi}},\ }\bibfield  {title} {\bibinfo {title}
  {Flexo-photovoltaic effect in mos2},\ }\href@noop {} {\bibfield  {journal}
  {\bibinfo  {journal} {Nature Nanotechnol.}\ }\textbf {\bibinfo {volume}
  {16}},\ \bibinfo {pages} {894} (\bibinfo {year} {2021})}\BibitemShut
  {NoStop}%
\bibitem [{\citenamefont {Chu}\ \emph {et~al.}(2015)\citenamefont {Chu} \emph
  {et~al.}}]{high9}%
  \BibitemOpen
  \bibfield  {author} {\bibinfo {author} {\bibfnamefont {K.}~\bibnamefont
  {Chu}} \emph {et~al.},\ }\bibfield  {title} {\bibinfo {title} {Enhancement of
  the anisotropic photocurrent in ferroelectric oxides by strain gradients},\
  }\href@noop {} {\bibfield  {journal} {\bibinfo  {journal} {Nature
  Nanotechnol.}\ }\textbf {\bibinfo {volume} {10}},\ \bibinfo {pages} {972}
  (\bibinfo {year} {2015})}\BibitemShut {NoStop}%
\bibitem [{\citenamefont {Yan}\ \emph {et~al.}(2020)\citenamefont {Yan} \emph
  {et~al.}}]{high11}%
  \BibitemOpen
  \bibfield  {author} {\bibinfo {author} {\bibfnamefont {X.}~\bibnamefont
  {Yan}} \emph {et~al.},\ }\bibfield  {title} {\bibinfo {title} {Continuously
  controllable photoconductance in freestanding bifeo3 by the macroscopic
  flexoelectric effect},\ }\href@noop {} {\bibfield  {journal} {\bibinfo
  {journal} {Nature Comm.}\ }\textbf {\bibinfo {volume} {11}},\ \bibinfo
  {pages} {2571} (\bibinfo {year} {2020})}\BibitemShut {NoStop}%
\bibitem [{\citenamefont {Wu}\ \emph {et~al.}(2021)\citenamefont {Wu},
  \citenamefont {Jiang}, \citenamefont {Lou}, \citenamefont {Zhang},
  \citenamefont {Song}, \citenamefont {Ning}, \citenamefont {Guo},
  \citenamefont {Pennycook}, \citenamefont {Dai},\ and\ \citenamefont
  {Wen}}]{high2}%
  \BibitemOpen
  \bibfield  {author} {\bibinfo {author} {\bibfnamefont {M.}~\bibnamefont
  {Wu}}, \bibinfo {author} {\bibfnamefont {Z.}~\bibnamefont {Jiang}}, \bibinfo
  {author} {\bibfnamefont {X.}~\bibnamefont {Lou}}, \bibinfo {author}
  {\bibfnamefont {F.}~\bibnamefont {Zhang}}, \bibinfo {author} {\bibfnamefont
  {D.}~\bibnamefont {Song}}, \bibinfo {author} {\bibfnamefont {S.}~\bibnamefont
  {Ning}}, \bibinfo {author} {\bibfnamefont {M.}~\bibnamefont {Guo}}, \bibinfo
  {author} {\bibfnamefont {S.~J.}\ \bibnamefont {Pennycook}}, \bibinfo {author}
  {\bibfnamefont {J.-Y.}\ \bibnamefont {Dai}},\ and\ \bibinfo {author}
  {\bibfnamefont {Z.}~\bibnamefont {Wen}},\ }\bibfield  {title} {\bibinfo
  {title} {Flexoelectric thin-film photodetectors},\ }\href@noop {} {\bibfield
  {journal} {\bibinfo  {journal} {Nano Lett.}\ }\textbf {\bibinfo {volume}
  {21}},\ \bibinfo {pages} {2946} (\bibinfo {year} {2021})}\BibitemShut
  {NoStop}%
\bibitem [{\citenamefont {Yun}\ \emph {et~al.}(2020)\citenamefont {Yun},
  \citenamefont {Song}, \citenamefont {Chu}, \citenamefont {Hwang},
  \citenamefont {Kim}, \citenamefont {Seo}, \citenamefont {Woo}, \citenamefont
  {Choi},\ and\ \citenamefont {Yang}}]{high1}%
  \BibitemOpen
  \bibfield  {author} {\bibinfo {author} {\bibfnamefont {S.}~\bibnamefont
  {Yun}}, \bibinfo {author} {\bibfnamefont {K.}~\bibnamefont {Song}}, \bibinfo
  {author} {\bibfnamefont {K.}~\bibnamefont {Chu}}, \bibinfo {author}
  {\bibfnamefont {S.-Y.}\ \bibnamefont {Hwang}}, \bibinfo {author}
  {\bibfnamefont {G.-Y.}\ \bibnamefont {Kim}}, \bibinfo {author} {\bibfnamefont
  {J.}~\bibnamefont {Seo}}, \bibinfo {author} {\bibfnamefont {C.-S.}\
  \bibnamefont {Woo}}, \bibinfo {author} {\bibfnamefont {S.-Y.}\ \bibnamefont
  {Choi}},\ and\ \bibinfo {author} {\bibfnamefont {C.-H.}\ \bibnamefont
  {Yang}},\ }\bibfield  {title} {\bibinfo {title} {Flexopiezoelectricity at
  ferroelastic domain walls in wo3 films},\ }\href@noop {} {\bibfield
  {journal} {\bibinfo  {journal} {Nature Comm.}\ }\textbf {\bibinfo {volume}
  {11}},\ \bibinfo {pages} {4898} (\bibinfo {year} {2020})}\BibitemShut
  {NoStop}%
\bibitem [{\citenamefont {Lee}\ \emph {et~al.}(2011)\citenamefont {Lee},
  \citenamefont {Yoon}, \citenamefont {Jang}, \citenamefont {Yoon},
  \citenamefont {Chung}, \citenamefont {Kim}, \citenamefont {Scott},\ and\
  \citenamefont {Noh}}]{high3}%
  \BibitemOpen
  \bibfield  {author} {\bibinfo {author} {\bibfnamefont {D.}~\bibnamefont
  {Lee}}, \bibinfo {author} {\bibfnamefont {A.}~\bibnamefont {Yoon}}, \bibinfo
  {author} {\bibfnamefont {S.}~\bibnamefont {Jang}}, \bibinfo {author}
  {\bibfnamefont {J.-G.}\ \bibnamefont {Yoon}}, \bibinfo {author}
  {\bibfnamefont {J.-S.}\ \bibnamefont {Chung}}, \bibinfo {author}
  {\bibfnamefont {M.}~\bibnamefont {Kim}}, \bibinfo {author} {\bibfnamefont
  {J.}~\bibnamefont {Scott}},\ and\ \bibinfo {author} {\bibfnamefont
  {T.}~\bibnamefont {Noh}},\ }\bibfield  {title} {\bibinfo {title} {{Giant
  Flexoelectric Effect in Ferroelectric Epitaxial Thin Films}},\ }\href@noop {}
  {\bibfield  {journal} {\bibinfo  {journal} {{Phys. Rev. Lett.}}\ }\textbf
  {\bibinfo {volume} {{107}}} (\bibinfo {year} {{2011}})}\BibitemShut {NoStop}%
\bibitem [{\citenamefont {Wang}\ \emph
  {et~al.}(2020{\natexlab{b}})\citenamefont {Wang}, \citenamefont {Jiang},
  \citenamefont {Wang}, \citenamefont {Stark}, \citenamefont {van Aken},
  \citenamefont {Mannhart},\ and\ \citenamefont {Boschker}}]{high7}%
  \BibitemOpen
  \bibfield  {author} {\bibinfo {author} {\bibfnamefont {H.}~\bibnamefont
  {Wang}}, \bibinfo {author} {\bibfnamefont {X.}~\bibnamefont {Jiang}},
  \bibinfo {author} {\bibfnamefont {Y.}~\bibnamefont {Wang}}, \bibinfo {author}
  {\bibfnamefont {R.~W.}\ \bibnamefont {Stark}}, \bibinfo {author}
  {\bibfnamefont {P.~A.}\ \bibnamefont {van Aken}}, \bibinfo {author}
  {\bibfnamefont {J.}~\bibnamefont {Mannhart}},\ and\ \bibinfo {author}
  {\bibfnamefont {H.}~\bibnamefont {Boschker}},\ }\bibfield  {title} {\bibinfo
  {title} {Direct observation of huge flexoelectric polarization around crack
  tips},\ }\href@noop {} {\bibfield  {journal} {\bibinfo  {journal} {Nano
  Lett.}\ }\textbf {\bibinfo {volume} {20}},\ \bibinfo {pages} {88} (\bibinfo
  {year} {2020}{\natexlab{b}})}\BibitemShut {NoStop}%
\bibitem [{\citenamefont {Gao}\ \emph {et~al.}(2018)\citenamefont {Gao},
  \citenamefont {Yang}, \citenamefont {Ishikawa}, \citenamefont {Li},
  \citenamefont {Feng}, \citenamefont {Kumamoto}, \citenamefont {Shibata},
  \citenamefont {Yu},\ and\ \citenamefont {Ikuhara}}]{high8}%
  \BibitemOpen
  \bibfield  {author} {\bibinfo {author} {\bibfnamefont {P.}~\bibnamefont
  {Gao}}, \bibinfo {author} {\bibfnamefont {S.}~\bibnamefont {Yang}}, \bibinfo
  {author} {\bibfnamefont {R.}~\bibnamefont {Ishikawa}}, \bibinfo {author}
  {\bibfnamefont {N.}~\bibnamefont {Li}}, \bibinfo {author} {\bibfnamefont
  {B.}~\bibnamefont {Feng}}, \bibinfo {author} {\bibfnamefont {A.}~\bibnamefont
  {Kumamoto}}, \bibinfo {author} {\bibfnamefont {N.}~\bibnamefont {Shibata}},
  \bibinfo {author} {\bibfnamefont {P.}~\bibnamefont {Yu}},\ and\ \bibinfo
  {author} {\bibfnamefont {Y.}~\bibnamefont {Ikuhara}},\ }\bibfield  {title}
  {\bibinfo {title} {Atomic-scale measurement of flexoelectric polarization at
  ${\mathrm{srtio}}_{3}$ dislocations},\ }\href@noop {} {\bibfield  {journal}
  {\bibinfo  {journal} {Phys. Rev. Lett.}\ }\textbf {\bibinfo {volume} {120}},\
  \bibinfo {pages} {267601} (\bibinfo {year} {2018})}\BibitemShut {NoStop}%
\bibitem [{\citenamefont {Lu}\ \emph {et~al.}(2012)\citenamefont {Lu},
  \citenamefont {Bark}, \citenamefont {de~los Ojos}, \citenamefont {Alcala},
  \citenamefont {Eom}, \citenamefont {Catalan},\ and\ \citenamefont
  {Gruverman}}]{revise4}%
  \BibitemOpen
  \bibfield  {author} {\bibinfo {author} {\bibfnamefont {H.}~\bibnamefont
  {Lu}}, \bibinfo {author} {\bibfnamefont {C.-W.}\ \bibnamefont {Bark}},
  \bibinfo {author} {\bibfnamefont {D.~E.}\ \bibnamefont {de~los Ojos}},
  \bibinfo {author} {\bibfnamefont {J.}~\bibnamefont {Alcala}}, \bibinfo
  {author} {\bibfnamefont {C.~B.}\ \bibnamefont {Eom}}, \bibinfo {author}
  {\bibfnamefont {G.}~\bibnamefont {Catalan}},\ and\ \bibinfo {author}
  {\bibfnamefont {A.}~\bibnamefont {Gruverman}},\ }\bibfield  {title} {\bibinfo
  {title} {Mechanical writing of ferroelectric polarization},\ }\href@noop {}
  {\bibfield  {journal} {\bibinfo  {journal} {Science}\ }\textbf {\bibinfo
  {volume} {336}},\ \bibinfo {pages} {59} (\bibinfo {year} {2012})}\BibitemShut
  {NoStop}%
\bibitem [{\citenamefont {Das}\ \emph {et~al.}(2019{\natexlab{a}})\citenamefont
  {Das}, \citenamefont {Wang}, \citenamefont {Paudel}, \citenamefont {Park},
  \citenamefont {Tsymbal}, \citenamefont {Chen}, \citenamefont {Lee},\ and\
  \citenamefont {W.}}]{revise5}%
  \BibitemOpen
  \bibfield  {author} {\bibinfo {author} {\bibfnamefont {S.}~\bibnamefont
  {Das}}, \bibinfo {author} {\bibfnamefont {B.}~\bibnamefont {Wang}}, \bibinfo
  {author} {\bibfnamefont {T.~R.}\ \bibnamefont {Paudel}}, \bibinfo {author}
  {\bibfnamefont {S.~M.}\ \bibnamefont {Park}}, \bibinfo {author}
  {\bibfnamefont {E.~Y.}\ \bibnamefont {Tsymbal}}, \bibinfo {author}
  {\bibfnamefont {L.-Q.}\ \bibnamefont {Chen}}, \bibinfo {author}
  {\bibfnamefont {D.}~\bibnamefont {Lee}},\ and\ \bibinfo {author}
  {\bibfnamefont {N.~T.}\ \bibnamefont {W.}},\ }\bibfield  {title} {\bibinfo
  {title} {Enhanced flexoelectricity at reduced dimensions revealed by
  mechanically tunable quantum tunnelling},\ }\href@noop {} {\bibfield
  {journal} {\bibinfo  {journal} {Nature Comm.}\ }\textbf {\bibinfo {volume}
  {10}},\ \bibinfo {pages} {537} (\bibinfo {year}
  {2019}{\natexlab{a}})}\BibitemShut {NoStop}%
\bibitem [{\citenamefont {Wang}\ \emph
  {et~al.}(2019{\natexlab{b}})\citenamefont {Wang}, \citenamefont {Song},
  \citenamefont {Shen}, \citenamefont {Huang}, \citenamefont {Li},
  \citenamefont {Ke},\ and\ \citenamefont {Shu}}]{high6}%
  \BibitemOpen
  \bibfield  {author} {\bibinfo {author} {\bibfnamefont {Z.}~\bibnamefont
  {Wang}}, \bibinfo {author} {\bibfnamefont {R.}~\bibnamefont {Song}}, \bibinfo
  {author} {\bibfnamefont {Z.}~\bibnamefont {Shen}}, \bibinfo {author}
  {\bibfnamefont {W.}~\bibnamefont {Huang}}, \bibinfo {author} {\bibfnamefont
  {C.}~\bibnamefont {Li}}, \bibinfo {author} {\bibfnamefont {S.}~\bibnamefont
  {Ke}},\ and\ \bibinfo {author} {\bibfnamefont {L.}~\bibnamefont {Shu}},\
  }\bibfield  {title} {\bibinfo {title} {Non-linear behavior of
  flexoelectricity},\ }\href@noop {} {\bibfield  {journal} {\bibinfo  {journal}
  {Appl. Phys. Lett.}\ }\textbf {\bibinfo {volume} {115}},\ \bibinfo {pages}
  {252905} (\bibinfo {year} {2019}{\natexlab{b}})}\BibitemShut {NoStop}%
\bibitem [{\citenamefont {Chu}\ and\ \citenamefont {Yang}(2017)}]{high4}%
  \BibitemOpen
  \bibfield  {author} {\bibinfo {author} {\bibfnamefont {K.}~\bibnamefont
  {Chu}}\ and\ \bibinfo {author} {\bibfnamefont {C.-H.}\ \bibnamefont {Yang}},\
  }\bibfield  {title} {\bibinfo {title} {Nonlinear flexoelectricity in
  noncentrosymmetric crystals},\ }\href@noop {} {\bibfield  {journal} {\bibinfo
   {journal} {Phys. Rev. B}\ }\textbf {\bibinfo {volume} {96}},\ \bibinfo
  {pages} {104102} (\bibinfo {year} {2017})}\BibitemShut {NoStop}%
\bibitem [{\citenamefont {Das}\ \emph {et~al.}(2019{\natexlab{b}})\citenamefont
  {Das}, \citenamefont {Wang}, \citenamefont {Paude}, \citenamefont {Park},
  \citenamefont {Tsymba}, \citenamefont {Chen}, \citenamefont {Lee},\ and\
  \citenamefont {Noh}}]{high5}%
  \BibitemOpen
  \bibfield  {author} {\bibinfo {author} {\bibfnamefont {S.}~\bibnamefont
  {Das}}, \bibinfo {author} {\bibfnamefont {B.}~\bibnamefont {Wang}}, \bibinfo
  {author} {\bibfnamefont {T.}~\bibnamefont {Paude}}, \bibinfo {author}
  {\bibfnamefont {S.~M.}\ \bibnamefont {Park}}, \bibinfo {author}
  {\bibfnamefont {E.}~\bibnamefont {Tsymba}}, \bibinfo {author} {\bibfnamefont
  {L.-Q.}\ \bibnamefont {Chen}}, \bibinfo {author} {\bibfnamefont
  {D.}~\bibnamefont {Lee}},\ and\ \bibinfo {author} {\bibfnamefont {T.~W.}\
  \bibnamefont {Noh}},\ }\bibfield  {title} {\bibinfo {title} {Enhanced
  flexoelectricity at reduced dimensions revealed by mechanically tunable
  quantum tunnelling},\ }\href@noop {} {\bibfield  {journal} {\bibinfo
  {journal} {Nature Comm.}\ }\textbf {\bibinfo {volume} {10}},\ \bibinfo
  {pages} {537} (\bibinfo {year} {2019}{\natexlab{b}})}\BibitemShut {NoStop}%
\bibitem [{\citenamefont {Hong}\ and\ \citenamefont
  {Vanderbilt}(2013)}]{hong3}%
  \BibitemOpen
  \bibfield  {author} {\bibinfo {author} {\bibfnamefont {J.}~\bibnamefont
  {Hong}}\ and\ \bibinfo {author} {\bibfnamefont {D.}~\bibnamefont
  {Vanderbilt}},\ }\bibfield  {title} {\bibinfo {title} {{First-principles
  theory and calculation of flexoelectricity}},\ }\href@noop {} {\bibfield
  {journal} {\bibinfo  {journal} {{Phys. Rev. B}}\ }\textbf {\bibinfo {volume}
  {{88}}},\ \bibinfo {pages} {{174107}} (\bibinfo {year} {{2013}})}\BibitemShut
  {NoStop}%
\bibitem [{\citenamefont {Zhang}\ and\ \citenamefont {Wei}(2017)}]{gbt1}%
  \BibitemOpen
  \bibfield  {author} {\bibinfo {author} {\bibfnamefont {D.-B.}\ \bibnamefont
  {Zhang}}\ and\ \bibinfo {author} {\bibfnamefont {S.-H.}\ \bibnamefont
  {Wei}},\ }\bibfield  {title} {\bibinfo {title} {Inhomogeneous strain-induced
  half-metallicity in bent zigzag graphene nanoribbons},\ }\href@noop {}
  {\bibfield  {journal} {\bibinfo  {journal} {Npj Comput. Mater.}\ }\textbf
  {\bibinfo {volume} {3}},\ \bibinfo {pages} {32} (\bibinfo {year}
  {2017})}\BibitemShut {NoStop}%
\bibitem [{\citenamefont {Yue}\ \emph {et~al.}(2017)\citenamefont {Yue},
  \citenamefont {Seifert}, \citenamefont {Chang},\ and\ \citenamefont
  {Zhang}}]{gbt2}%
  \BibitemOpen
  \bibfield  {author} {\bibinfo {author} {\bibfnamefont {L.}~\bibnamefont
  {Yue}}, \bibinfo {author} {\bibfnamefont {G.}~\bibnamefont {Seifert}},
  \bibinfo {author} {\bibfnamefont {K.}~\bibnamefont {Chang}},\ and\ \bibinfo
  {author} {\bibfnamefont {D.-B.}\ \bibnamefont {Zhang}},\ }\bibfield  {title}
  {\bibinfo {title} {Effective zeeman splitting in bent lateral heterojunctions
  of graphene and hexagonal boron nitride: A new mechanism towards
  half-metallicity},\ }\href@noop {} {\bibfield  {journal} {\bibinfo  {journal}
  {Phys. Rev. B}\ }\textbf {\bibinfo {volume} {96}},\ \bibinfo {pages} {201403}
  (\bibinfo {year} {2017})}\BibitemShut {NoStop}%
\bibitem [{\citenamefont {Tang}\ \emph {et~al.}(2023)\citenamefont {Tang},
  \citenamefont {Wang}, \citenamefont {Chang},\ and\ \citenamefont
  {Zhang}}]{gbt5}%
  \BibitemOpen
  \bibfield  {author} {\bibinfo {author} {\bibfnamefont {J.-K.}\ \bibnamefont
  {Tang}}, \bibinfo {author} {\bibfnamefont {Y.-X.}\ \bibnamefont {Wang}},
  \bibinfo {author} {\bibfnamefont {K.}~\bibnamefont {Chang}},\ and\ \bibinfo
  {author} {\bibfnamefont {D.-B.}\ \bibnamefont {Zhang}},\ }\bibfield  {title}
  {\bibinfo {title} {Polarization due to emergent polarity in elemental
  semiconductor thinfilms under bending},\ }\href@noop {} {\bibfield  {journal}
  {\bibinfo  {journal} {J. Phys.: Condens. Matter}\ }\textbf {\bibinfo {volume}
  {35}},\ \bibinfo {pages} {015501} (\bibinfo {year} {2023})}\BibitemShut
  {NoStop}%
\bibitem [{\citenamefont {Codony}\ \emph {et~al.}(2021)\citenamefont {Codony},
  \citenamefont {Arias},\ and\ \citenamefont {Suryanarayana}}]{rotate}%
  \BibitemOpen
  \bibfield  {author} {\bibinfo {author} {\bibfnamefont {D.}~\bibnamefont
  {Codony}}, \bibinfo {author} {\bibfnamefont {I.}~\bibnamefont {Arias}},\ and\
  \bibinfo {author} {\bibfnamefont {P.}~\bibnamefont {Suryanarayana}},\
  }\bibfield  {title} {\bibinfo {title} {Transversal flexoelectric coefficient
  for nanostructures at finite deformations from first principles},\
  }\href@noop {} {\bibfield  {journal} {\bibinfo  {journal} {Phys. Rev.
  Mater.}\ }\textbf {\bibinfo {volume} {5}},\ \bibinfo {pages} {L030801}
  (\bibinfo {year} {2021})}\BibitemShut {NoStop}%
\bibitem [{\citenamefont {White}\ \emph {et~al.}(1993)\citenamefont {White},
  \citenamefont {Robertson},\ and\ \citenamefont {Mintmire}}]{rotate2}%
  \BibitemOpen
  \bibfield  {author} {\bibinfo {author} {\bibfnamefont {C.~T.}\ \bibnamefont
  {White}}, \bibinfo {author} {\bibfnamefont {D.~H.}\ \bibnamefont
  {Robertson}},\ and\ \bibinfo {author} {\bibfnamefont {J.}~\bibnamefont
  {Mintmire}},\ }\bibfield  {title} {\bibinfo {title} {Helical and rotational
  symmetries of nanoscale graphitic tubules},\ }\href@noop {} {\bibfield
  {journal} {\bibinfo  {journal} {Phys. Rev. B}\ }\textbf {\bibinfo {volume}
  {47}},\ \bibinfo {pages} {5485} (\bibinfo {year} {1993})}\BibitemShut
  {NoStop}%
\bibitem [{\citenamefont {Popov}(2004)}]{rotate3}%
  \BibitemOpen
  \bibfield  {author} {\bibinfo {author} {\bibfnamefont {V.~N.}\ \bibnamefont
  {Popov}},\ }\bibfield  {title} {\bibinfo {title} {Curvature effects on the
  structural, electronic and optical properties of isolated single-walled
  carbon nanotubes within a symmetry-adapted non-orthogonal tight-binding
  model},\ }\href@noop {} {\bibfield  {journal} {\bibinfo  {journal} {New J.
  Phys.}\ }\textbf {\bibinfo {volume} {6}},\ \bibinfo {pages} {17} (\bibinfo
  {year} {2004})}\BibitemShut {NoStop}%
\bibitem [{\citenamefont {Allen}(2007)}]{rotate4}%
  \BibitemOpen
  \bibfield  {author} {\bibinfo {author} {\bibfnamefont {P.~B.}\ \bibnamefont
  {Allen}},\ }\bibfield  {title} {\bibinfo {title} {Nanocrystalline nanowires:
  Iii electrons},\ }\href@noop {} {\bibfield  {journal} {\bibinfo  {journal}
  {Nano Lett.}\ }\textbf {\bibinfo {volume} {7}},\ \bibinfo {pages} {1220}
  (\bibinfo {year} {2007})}\BibitemShut {NoStop}%
\bibitem [{\citenamefont {Stengel}(2013{\natexlab{b}})}]{Stengel1}%
  \BibitemOpen
  \bibfield  {author} {\bibinfo {author} {\bibfnamefont {M.}~\bibnamefont
  {Stengel}},\ }\bibfield  {title} {\bibinfo {title} {Microscopic response to
  inhomogeneous deformations in curvilinear coordinates},\ }\href@noop {}
  {\bibfield  {journal} {\bibinfo  {journal} {Nature Commun.}\ }\textbf
  {\bibinfo {volume} {4}},\ \bibinfo {pages} {2693} (\bibinfo {year}
  {2013}{\natexlab{b}})}\BibitemShut {NoStop}%
\bibitem [{\citenamefont {Porezag}\ \emph {et~al.}(1995)\citenamefont
  {Porezag}, \citenamefont {Frauenheim}, \citenamefont {K\"ohler},
  \citenamefont {Seifert},\ and\ \citenamefont {Kaschner}}]{dftb1}%
  \BibitemOpen
  \bibfield  {author} {\bibinfo {author} {\bibfnamefont {D.}~\bibnamefont
  {Porezag}}, \bibinfo {author} {\bibfnamefont {T.}~\bibnamefont {Frauenheim}},
  \bibinfo {author} {\bibfnamefont {T.}~\bibnamefont {K\"ohler}}, \bibinfo
  {author} {\bibfnamefont {G.}~\bibnamefont {Seifert}},\ and\ \bibinfo {author}
  {\bibfnamefont {R.}~\bibnamefont {Kaschner}},\ }\bibfield  {title} {\bibinfo
  {title} {Construction of tight-binding-like potentials on the basis of
  density-functional theory: Application to carbon},\ }\href
  {https://doi.org/10.1103/PhysRevB.51.12947} {\bibfield  {journal} {\bibinfo
  {journal} {Phys. Rev. B}\ }\textbf {\bibinfo {volume} {51}},\ \bibinfo
  {pages} {12947} (\bibinfo {year} {1995})}\BibitemShut {NoStop}%
\bibitem [{\citenamefont {Rurali}\ and\ \citenamefont
  {Hernandez}(2003)}]{dftb2}%
  \BibitemOpen
  \bibfield  {author} {\bibinfo {author} {\bibfnamefont {R.}~\bibnamefont
  {Rurali}}\ and\ \bibinfo {author} {\bibfnamefont {E.}~\bibnamefont
  {Hernandez}},\ }\bibfield  {title} {\bibinfo {title} {Trocadero: a
  multiple-algorithm multiple-model atomistic simulation program},\ }\href
  {https://doi.org/https://doi.org/10.1016/S0927-0256(03)00100-9} {\bibfield
  {journal} {\bibinfo  {journal} {Comput. Mater. Sci.}\ }\textbf {\bibinfo
  {volume} {28}},\ \bibinfo {pages} {85} (\bibinfo {year} {2003})}\BibitemShut
  {NoStop}%
\bibitem [{\citenamefont {Elstner}\ \emph {et~al.}(1998)\citenamefont
  {Elstner}, \citenamefont {Porezag}, \citenamefont {Jungnickel}, \citenamefont
  {Elsner}, \citenamefont {Haugk}, \citenamefont {Frauenheim}, \citenamefont
  {Suhai},\ and\ \citenamefont {Seifert}}]{dftb3}%
  \BibitemOpen
  \bibfield  {author} {\bibinfo {author} {\bibfnamefont {M.}~\bibnamefont
  {Elstner}}, \bibinfo {author} {\bibfnamefont {D.}~\bibnamefont {Porezag}},
  \bibinfo {author} {\bibfnamefont {G.}~\bibnamefont {Jungnickel}}, \bibinfo
  {author} {\bibfnamefont {J.}~\bibnamefont {Elsner}}, \bibinfo {author}
  {\bibfnamefont {M.}~\bibnamefont {Haugk}}, \bibinfo {author} {\bibfnamefont
  {T.}~\bibnamefont {Frauenheim}}, \bibinfo {author} {\bibfnamefont
  {S.}~\bibnamefont {Suhai}},\ and\ \bibinfo {author} {\bibfnamefont
  {G.}~\bibnamefont {Seifert}},\ }\bibfield  {title} {\bibinfo {title}
  {Self-consistent-charge density-functional tight-binding method for
  simulations of complex materials properties},\ }\href@noop {} {\bibfield
  {journal} {\bibinfo  {journal} {Phys. Rev. B}\ }\textbf {\bibinfo {volume}
  {58}},\ \bibinfo {pages} {7260} (\bibinfo {year} {1998})}\BibitemShut
  {NoStop}%
\bibitem [{\citenamefont {Bl\"ochl}(1994)}]{vasp1}%
  \BibitemOpen
  \bibfield  {author} {\bibinfo {author} {\bibfnamefont {P.~E.}\ \bibnamefont
  {Bl\"ochl}},\ }\bibfield  {title} {\bibinfo {title} {Projector augmented-wave
  method},\ }\href {https://doi.org/10.1103/PhysRevB.50.17953} {\bibfield
  {journal} {\bibinfo  {journal} {Phys. Rev. B}\ }\textbf {\bibinfo {volume}
  {50}},\ \bibinfo {pages} {17953} (\bibinfo {year} {1994})}\BibitemShut
  {NoStop}%
\bibitem [{\citenamefont {Kresse}\ and\ \citenamefont {Joubert}(1999)}]{vasp2}%
  \BibitemOpen
  \bibfield  {author} {\bibinfo {author} {\bibfnamefont {G.}~\bibnamefont
  {Kresse}}\ and\ \bibinfo {author} {\bibfnamefont {D.}~\bibnamefont
  {Joubert}},\ }\bibfield  {title} {\bibinfo {title} {From ultrasoft
  pseudopotentials to the projector augmented-wave method},\ }\href
  {https://doi.org/10.1103/PhysRevB.59.1758} {\bibfield  {journal} {\bibinfo
  {journal} {Phys. Rev. B}\ }\textbf {\bibinfo {volume} {59}},\ \bibinfo
  {pages} {1758} (\bibinfo {year} {1999})}\BibitemShut {NoStop}%
\bibitem [{\citenamefont {Perdew}\ \emph {et~al.}(1996)\citenamefont {Perdew},
  \citenamefont {Burke},\ and\ \citenamefont {Ernzerhof}}]{vasp3}%
  \BibitemOpen
  \bibfield  {author} {\bibinfo {author} {\bibfnamefont {J.~P.}\ \bibnamefont
  {Perdew}}, \bibinfo {author} {\bibfnamefont {K.}~\bibnamefont {Burke}},\ and\
  \bibinfo {author} {\bibfnamefont {M.}~\bibnamefont {Ernzerhof}},\ }\bibfield
  {title} {\bibinfo {title} {Generalized gradient approximation made simple},\
  }\href {https://doi.org/10.1103/PhysRevLett.77.3865} {\bibfield  {journal}
  {\bibinfo  {journal} {Phys. Rev. Lett.}\ }\textbf {\bibinfo {volume} {77}},\
  \bibinfo {pages} {3865} (\bibinfo {year} {1996})}\BibitemShut {NoStop}%
\bibitem [{\citenamefont {Kresse}\ and\ \citenamefont
  {Furthm\"uller}(1996)}]{vasp}%
  \BibitemOpen
  \bibfield  {author} {\bibinfo {author} {\bibfnamefont {G.}~\bibnamefont
  {Kresse}}\ and\ \bibinfo {author} {\bibfnamefont {J.}~\bibnamefont
  {Furthm\"uller}},\ }\bibfield  {title} {\bibinfo {title} {Efficient iterative
  schemes for ab initio total-energy calculations using a plane-wave basis
  set},\ }\href {https://doi.org/10.1103/PhysRevB.54.11169} {\bibfield
  {journal} {\bibinfo  {journal} {Phys. Rev. B}\ }\textbf {\bibinfo {volume}
  {54}},\ \bibinfo {pages} {11169} (\bibinfo {year} {1996})}\BibitemShut
  {NoStop}%
\bibitem [{\citenamefont {Scanlon}\ \emph {et~al.}(2013)\citenamefont {Scanlon}
  \emph {et~al.}}]{typeii1}%
  \BibitemOpen
  \bibfield  {author} {\bibinfo {author} {\bibfnamefont {D.}~\bibnamefont
  {Scanlon}} \emph {et~al.},\ }\bibfield  {title} {\bibinfo {title} {Band
  alignment of rutile and anatase tio2},\ }\href@noop {} {\bibfield  {journal}
  {\bibinfo  {journal} {Nature Mater.}\ }\textbf {\bibinfo {volume} {12}},\
  \bibinfo {pages} {798} (\bibinfo {year} {2013})}\BibitemShut {NoStop}%
\bibitem [{\citenamefont {Rivera}\ \emph {et~al.}(2015)\citenamefont {Rivera}
  \emph {et~al.}}]{typeii2}%
  \BibitemOpen
  \bibfield  {author} {\bibinfo {author} {\bibfnamefont {P.}~\bibnamefont
  {Rivera}} \emph {et~al.},\ }\bibfield  {title} {\bibinfo {title} {Observation
  of long-lived interlayer excitons in monolayer mose2-wse2 heterostructures},\
  }\href@noop {} {\bibfield  {journal} {\bibinfo  {journal} {Nature Comm.}\
  }\textbf {\bibinfo {volume} {6}},\ \bibinfo {pages} {6242} (\bibinfo {year}
  {2015})}\BibitemShut {NoStop}%
\bibitem [{\citenamefont {Gong}\ \emph {et~al.}(2014)\citenamefont {Gong} \emph
  {et~al.}}]{typeii3}%
  \BibitemOpen
  \bibfield  {author} {\bibinfo {author} {\bibfnamefont {Y.}~\bibnamefont
  {Gong}} \emph {et~al.},\ }\bibfield  {title} {\bibinfo {title} {Vertical and
  in-plane heterostructures from ws2/mos2 monolayers},\ }\href@noop {}
  {\bibfield  {journal} {\bibinfo  {journal} {Nature Mater.}\ }\textbf
  {\bibinfo {volume} {13}},\ \bibinfo {pages} {1135} (\bibinfo {year}
  {2014})}\BibitemShut {NoStop}%
\bibitem [{\citenamefont {Tagantsev}(1986)}]{Tagantsev}%
  \BibitemOpen
  \bibfield  {author} {\bibinfo {author} {\bibfnamefont {A.~K.}\ \bibnamefont
  {Tagantsev}},\ }\bibfield  {title} {\bibinfo {title} {{Piezoelectricity and
  flexoelectricity in crystalline dielectrics}},\ }\href@noop {} {\bibfield
  {journal} {\bibinfo  {journal} {{Phys. Rev. B}}\ }\textbf {\bibinfo {volume}
  {{34}}},\ \bibinfo {pages} {{5883}} (\bibinfo {year} {{1986}})}\BibitemShut
  {NoStop}%
\bibitem [{\citenamefont {Resta}(2010)}]{Resta1}%
  \BibitemOpen
  \bibfield  {author} {\bibinfo {author} {\bibfnamefont {R.}~\bibnamefont
  {Resta}},\ }\bibfield  {title} {\bibinfo {title} {{Towards a Bulk Theory of
  Flexoelectricity}},\ }\href@noop {} {\bibfield  {journal} {\bibinfo
  {journal} {{Phys. Rev. Lett. }}\ }\textbf {\bibinfo {volume} {{105}}},\
  \bibinfo {pages} {{127601}} (\bibinfo {year} {{2010}})}\BibitemShut {NoStop}%
\bibitem [{\citenamefont {Hong}\ and\ \citenamefont
  {Vanderbilt}(2011)}]{hong2}%
  \BibitemOpen
  \bibfield  {author} {\bibinfo {author} {\bibfnamefont {J.}~\bibnamefont
  {Hong}}\ and\ \bibinfo {author} {\bibfnamefont {D.}~\bibnamefont
  {Vanderbilt}},\ }\bibfield  {title} {\bibinfo {title} {{First-principles
  theory of frozen-ion flexoelectricity}},\ }\href@noop {} {\bibfield
  {journal} {\bibinfo  {journal} {{Phys. Rev. B}}\ }\textbf {\bibinfo {volume}
  {{84}}},\ \bibinfo {pages} {{180101}} (\bibinfo {year} {{2011}})}\BibitemShut
  {NoStop}%
\bibitem [{\citenamefont {Stengel}(2014)}]{Stengel2}%
  \BibitemOpen
  \bibfield  {author} {\bibinfo {author} {\bibfnamefont {M.}~\bibnamefont
  {Stengel}},\ }\bibfield  {title} {\bibinfo {title} {Surface control of
  flexoelectricity},\ }\href@noop {} {\bibfield  {journal} {\bibinfo  {journal}
  {Phys. Rev. B}\ }\textbf {\bibinfo {volume} {90}},\ \bibinfo {pages} {201112}
  (\bibinfo {year} {2014})}\BibitemShut {NoStop}%
\bibitem [{\citenamefont {Mizzi}\ and\ \citenamefont
  {Marks}(2021)}]{surfacejap}%
  \BibitemOpen
  \bibfield  {author} {\bibinfo {author} {\bibfnamefont {C.~A.}\ \bibnamefont
  {Mizzi}}\ and\ \bibinfo {author} {\bibfnamefont {L.~D.}\ \bibnamefont
  {Marks}},\ }\bibfield  {title} {\bibinfo {title} {{The role of surfaces in
  flexoelectricity}},\ }\href@noop {} {\bibfield  {journal} {\bibinfo
  {journal} {Journal of Applied Physics}\ }\textbf {\bibinfo {volume} {129}},\
  \bibinfo {pages} {224102} (\bibinfo {year} {2021})}\BibitemShut {NoStop}%
\bibitem [{\citenamefont {Zhang}\ \emph {et~al.}(2016)\citenamefont {Zhang},
  \citenamefont {Tersoff}, \citenamefont {Xu}, \citenamefont {Chen},
  \citenamefont {Zhang}, \citenamefont {Zhang}, \citenamefont {Yang},
  \citenamefont {Lee}, \citenamefont {Tu}, \citenamefont {Li},\ and\
  \citenamefont {Lu}}]{strains}%
  \BibitemOpen
  \bibfield  {author} {\bibinfo {author} {\bibfnamefont {H.}~\bibnamefont
  {Zhang}}, \bibinfo {author} {\bibfnamefont {J.}~\bibnamefont {Tersoff}},
  \bibinfo {author} {\bibfnamefont {S.}~\bibnamefont {Xu}}, \bibinfo {author}
  {\bibfnamefont {H.}~\bibnamefont {Chen}}, \bibinfo {author} {\bibfnamefont
  {Q.}~\bibnamefont {Zhang}}, \bibinfo {author} {\bibfnamefont
  {K.}~\bibnamefont {Zhang}}, \bibinfo {author} {\bibfnamefont
  {Y.}~\bibnamefont {Yang}}, \bibinfo {author} {\bibfnamefont {C.-S.}\
  \bibnamefont {Lee}}, \bibinfo {author} {\bibfnamefont {K.-N.}\ \bibnamefont
  {Tu}}, \bibinfo {author} {\bibfnamefont {J.}~\bibnamefont {Li}},\ and\
  \bibinfo {author} {\bibfnamefont {Y.}~\bibnamefont {Lu}},\ }\bibfield
  {title} {\bibinfo {title} {Approaching the ideal elastic strain limit in
  silicon nanowires},\ }\href@noop {} {\bibfield  {journal} {\bibinfo
  {journal} {Sci. Adv.}\ }\textbf {\bibinfo {volume} {2}},\ \bibinfo {pages}
  {e1501382} (\bibinfo {year} {2016})}\BibitemShut {NoStop}%
\bibitem [{\citenamefont {Heremans}\ \emph {et~al.}(2008)\citenamefont
  {Heremans}, \citenamefont {Jovovic}, \citenamefont {Toberer}, \citenamefont
  {Saramat}, \citenamefont {Kurosaki}, \citenamefont {Charoenphakdee},
  \citenamefont {Yamanaka},\ and\ \citenamefont {Snyder}}]{pbte}%
  \BibitemOpen
  \bibfield  {author} {\bibinfo {author} {\bibfnamefont {J.~P.}\ \bibnamefont
  {Heremans}}, \bibinfo {author} {\bibfnamefont {V.}~\bibnamefont {Jovovic}},
  \bibinfo {author} {\bibfnamefont {E.~S.}\ \bibnamefont {Toberer}}, \bibinfo
  {author} {\bibfnamefont {A.}~\bibnamefont {Saramat}}, \bibinfo {author}
  {\bibfnamefont {K.}~\bibnamefont {Kurosaki}}, \bibinfo {author}
  {\bibfnamefont {A.}~\bibnamefont {Charoenphakdee}}, \bibinfo {author}
  {\bibfnamefont {S.}~\bibnamefont {Yamanaka}},\ and\ \bibinfo {author}
  {\bibfnamefont {G.~J.}\ \bibnamefont {Snyder}},\ }\bibfield  {title}
  {\bibinfo {title} {Enhancement of thermoelectric efficiency in pbte by
  distortion of the electronic density of states},\ }\href@noop {} {\bibfield
  {journal} {\bibinfo  {journal} {Science}\ }\textbf {\bibinfo {volume}
  {321}},\ \bibinfo {pages} {554} (\bibinfo {year} {2008})}\BibitemShut
  {NoStop}%
\bibitem [{\citenamefont {Skelton}\ \emph {et~al.}(2014)\citenamefont
  {Skelton}, \citenamefont {Parker}, \citenamefont {Togo}, \citenamefont
  {Tanaka},\ and\ \citenamefont {Walsh}}]{pbx}%
  \BibitemOpen
  \bibfield  {author} {\bibinfo {author} {\bibfnamefont {J.~M.}\ \bibnamefont
  {Skelton}}, \bibinfo {author} {\bibfnamefont {S.~C.}\ \bibnamefont {Parker}},
  \bibinfo {author} {\bibfnamefont {A.}~\bibnamefont {Togo}}, \bibinfo {author}
  {\bibfnamefont {I.}~\bibnamefont {Tanaka}},\ and\ \bibinfo {author}
  {\bibfnamefont {A.}~\bibnamefont {Walsh}},\ }\bibfield  {title} {\bibinfo
  {title} {Thermal physics of the lead chalcogenides pbs, pbse, and pbte from
  first principles},\ }\href@noop {} {\bibfield  {journal} {\bibinfo  {journal}
  {Phys. Rev. B}\ }\textbf {\bibinfo {volume} {89}},\ \bibinfo {pages} {205203}
  (\bibinfo {year} {2014})}\BibitemShut {NoStop}%
\bibitem [{\citenamefont {Springholz}\ and\ \citenamefont {Bauer}()}]{viiv}%
  \BibitemOpen
  \bibfield  {author} {\bibinfo {author} {\bibfnamefont {G.}~\bibnamefont
  {Springholz}}\ and\ \bibinfo {author} {\bibfnamefont {G.}~\bibnamefont
  {Bauer}},\ }\bibfield  {title} {\bibinfo {title} {Semiconductors, iv-vi,
  wiley encycl. electr. electron. eng.},\ }\href@noop {} {\bibinfo  {journal}
  {(Wiley, 2014), pp.1-16.}\ }\BibitemShut {NoStop}%
\bibitem [{\citenamefont {Singh}()}]{singh}%
  \BibitemOpen
\bibfield  {journal} {  }\bibfield  {author} {\bibinfo {author} {\bibfnamefont
  {J.}~\bibnamefont {Singh}},\ }\bibfield  {title} {\bibinfo {title} {Physics
  of semiconductors and their heterostructures},\ }\href@noop {} {\bibinfo
  {journal} {(McGraw-Hill College, 1992)}\ }\BibitemShut {NoStop}%
\bibitem [{\citenamefont {Harrison}()}]{harrison}%
  \BibitemOpen
\bibfield  {journal} {  }\bibfield  {author} {\bibinfo {author} {\bibfnamefont
  {W.~A.}\ \bibnamefont {Harrison}},\ }\bibfield  {title} {\bibinfo {title}
  {Electronic structure and the properties of solids: The physics of the
  chemical bond},\ }\href@noop {} {\bibinfo  {journal} {(Dover Publications,
  1981)}\ }\BibitemShut {NoStop}%
\end{thebibliography}%

\end{document}